# Search for Sources of Ultrahigh Energy Cosmic Rays

A.A. Mikhailov

*Institute of Cosmophysical Research and Aeronomy, 31 Lenin Ave., 677891 Yakutsk, Russia*

**Abstract**

The connection between the arrival directions of ultrahigh energy cosmic rays by using the EAS array data and point galactic sources of radio - and gamma - radiations, pulsars, is sought. At the mean particle energy of $\sim 10^{19}$ eV the correlation between the particle arrival directions and pulsars located along the magnetic field lines has been found. According to our estimations the chance probability is equal to $2 \cdot 10^{-4}$. The observed particle flux from 16 pulsars inside a circle of radius $< 6°$ exceeds the background by $6.3\sigma$ ($p < 10^{-10}$), from PSR 2351+61 - $5\sigma$. A group of 9 pulsars has been found from the direction of which the particle flux in the region $10°\times 60°$ is in excess of the expected one by $4.4\sigma$. The obtained results are discussed.

In (Mikhailov,1988) it was shown that by Yakutsk EAS (extensive air shower) array data cosmic rays at energies $\sim 10^{19}$ eV correlated with pulsars from the direction along the magnetic field lines. In our paper we considered data for 1974-1995 and carried out the further analysis.

576 showers with energies $(0.8-4)\cdot 10^{19}$ eV at the zenith arrival angles $< 60°$ and axes located inside of the array perimeter are considered. An definition accuracy the primary particle energy generating EAS is $\sim 30\%$, and for the arrival angle is $\sim 3°$. Data on the pulsars are taken from Line & Graham-Smith, 1990.

The angular distance between the arrival direction of each shower and each pulsar was calculated. If this distance is smaller than the definite radius r from the given pulsar then the shower was considered to be related to this pulsar. Each shower was considered only one time and only to the nearest pulsar.

The observed events (circles, they are connected by a solid line for the comparison) and expected (dashes lines) events at the isotropic cosmic ray distribution for separate arrival directions inside a circle of the radius $< r$ (integral characteristics) are presented in Fig.1: (a) - from the whole visible part of the celestial sphere, $\delta > 2°$ ($\delta$ is a declination); (b) - from the direction of the galactic halo - $|b| > 30°$ (b is a galactic latitude); (c) - from the direction of the Galaxy anticenter - $|b| < 30°$ и $120° < l < 240°$ (l is a galactic longitude); (d) - from the side along the magnetic field lines - a circle of the radius $< 45°$ around the direction $b = 0°, l = 90°$.

The Pearson $\chi^2$ - test for the first three cases (Fig.1,a-c) shows that the observed events do not contradict to the expected ones ($p \sim 0.9$). The Monte-Carlo method gives the same result: for example, at radius $r < 4°$, $6°$, $8°$ the probabilities are $p > 0.5$. The $\chi^2$ - test for the shower arrival from the direction along the magnetic field lines (Fig.1,d) shows that the observed and expected events are not in agreement ($p < 0.5$). At $r < 6°$ where the deviation is maximum the number of observed events is $n_{obs} = 157$ and of expected ones is $n_{exp} = 130.1$. According to the Monte Carlo method, the chance probability is $p \sim 2 \cdot 10^{-4}$. At $r < 4°, 8°, 10°$ the probabilities p are $0.1; 3 \cdot 10^{-4}; 0.01$, respectively.

The distribution of 16 pulsars on the celestial sphere for which the number of events at $r < 6°$ exceeds the background by $(1 - 5)\sigma$ is presented in Fig.2. For example, for the pulsar PSR 2351+61 19 events are observed (if to assume that all the events at $r < 6°$ are related to this pulsar) instead of 6.3 expected ones (the exceeding is $\sim 5\sigma$).

Fig.3 demonstrates the distribution of the observed and expected number of events from 16 pulsars. The $\chi^2$ - test shows that they do not agree with each other ($p < 0.001$). From these pulsars for example at $r < 6°$ the exceeding of $n_{obs} = 98$ above $n_{exp} = 52.4$ is $6.3\sigma$ ($p < 10^{-10}$). At $r < 4°, 8°$ the exceeding of the observed events above the expected ones is $4.9\sigma$ and $4.6\sigma$, respectively ($p < 10^{-6}$).

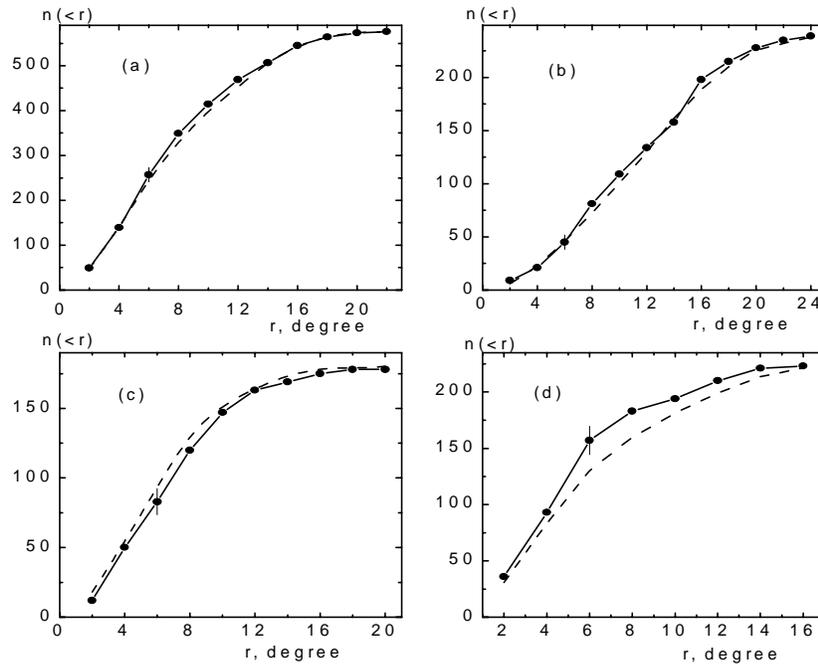

Fig1. Observed (circles) and expected (dashed lines) events of EAS inside the circle < r: (a) - from the hemisphere, (b) - from the halo, ( c ) - from the anticenter, (d) - from along the magnetic field lines

For check of a method of the analysis we determined a flux of particles from chance point with coordinates $\delta = 40°$, $\alpha = 150°$ not coincides with pulsars and located in region, where the density of pulsars is small (fig.2). On fig.4 the distribution of observed events < r from this point and of expected in case of an isotropy is shown. As it is shown from fig.4, the number of observed events is less expected in all considered angle. For r < 12°, 14°, 16° numbers of observed events less $3\sigma$ than expected. ($3.8\sigma$, $3.8\sigma$, $3.5\sigma$ respectively). From here we can to conclude, that where there is no pulsars - a flux of particles below expected in case of an isotropy. As seen from Fig.2, the main part (a group of 9 pulsars) is located in the region of the celestial sphere with $\delta \sim 60°$, $\alpha \sim 340° - 30°$ ($\alpha$ - right ascension). From 9 pulsars at r < 4°, 6°, 8° the exceeding of the observed events above the expected ones is $6.2\sigma$ ($n_{obs} = 32$, $n_{exp} = 11.2$), $5.0\sigma$ and $3.5\sigma$ respectively.

If we assume, that each shower relates is to all pulsars(not only nearest), we find in addition a group of 10 pulsars, located in the region of a maximum concentration of pulsars(fig.2, $\sigma \sim 25°$, $\alpha \sim 290°$). The flux of particles from direction of this group pulsars is small, but at r < 4°, 6° exceedings of observed events above expected is more

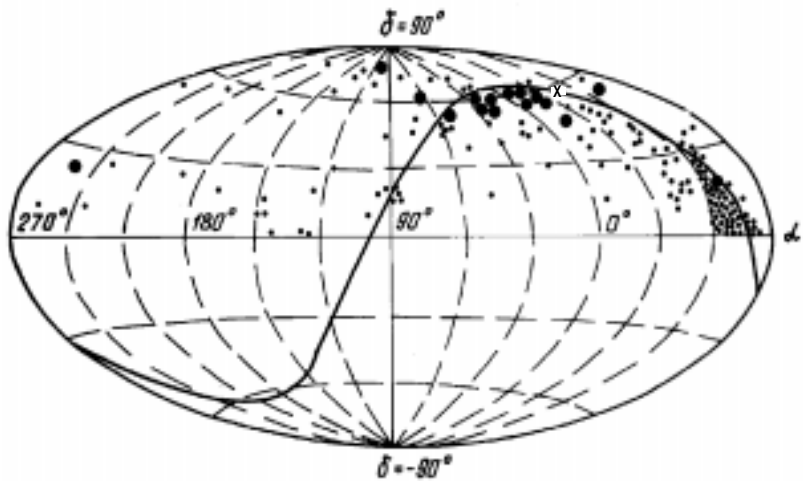

Fig.2. The distribution of the pulsars: ● - 16 pulsars, • - the ordinary pulsars. x - b = 0°, l = 90°. Solid line is the Galactic Plane

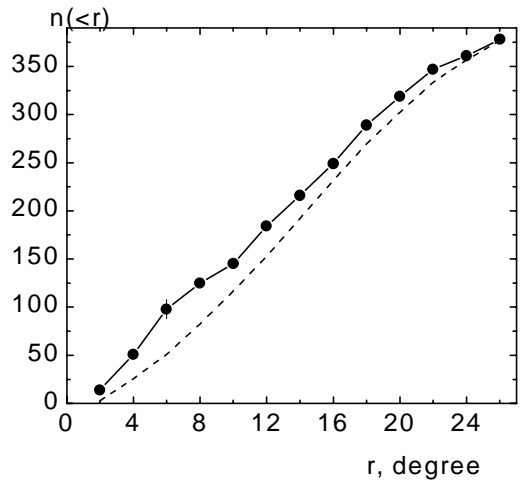

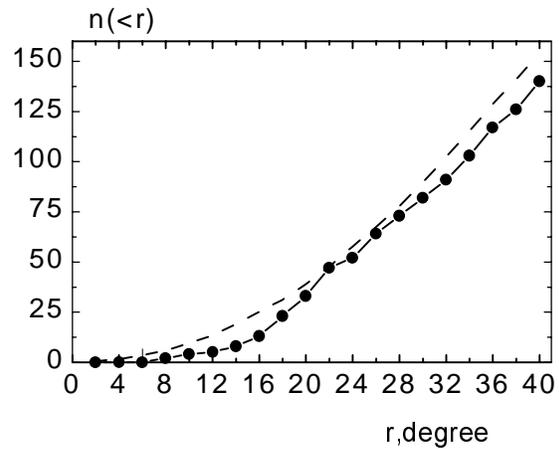

Fig.3. Observed events of EAS from 16 pulsars(circles) and expected (dashed lines)

Fig.4. Observed events of EAS from chance point not coincides with pulsars(circles) and expected (dashed lines)

$2\sigma$ - $4.5\sigma$($n_{obs}$ = 10, $n_{exp}$ = 2.7), $3.4\sigma$ respectively.

Fig.5 shows the distribution of showers in the equatorial coordinate by $10° \times 30°$ region as the mean square deviation $\sigma$ of the observed number of showers from the expected ones in the case of isotropy. It is seen that three maxima shower arrival direction are observed: region 1 - $\delta \sim 60°$, $\alpha \sim 330°$ - $30°$; region 2 - $\delta = 20°$ - $30°$, $\alpha = 270°$ - $300°$, region 3 - $\delta = 0°$ - $10°$, $\alpha = 180°$ - $270°$. If located side by side regions with deviations $> 2\sigma$ to combine then we'll obtain the excess of the observed particles above the expected one by $3.4\sigma$ ($n_{obs}$ = 41, $n_{exp}$ = 24.1) in region 1 and $4.5\sigma$ ($n_{obs}$ = 14, $n_{exp}$ = 4.5) in the region 3. Regions 1, 2 are in agreement with coordinates of 9,10 pulsars respectively, region 2 is in the high latitudes of the central part of Galaxy.

If we consider the particle distribution in the galactic coordinate (Fig.6) then the exceeding of particle flux by $4.4\sigma$ ($n_{obs}$ = 51, $n_{exp}$ = 27.7) is observed at b = $-10°$ - $0°$, l = $90°$ - $150°$ where 9 pulsars are located (Fig.2 and see Table). The deviation of the observed number of particles ($n_{obs}$ = 73) in comparison with the expected one ($n_{exp}$ = 51.1) in the case of the isotropy is maximum 3.1σ in the galactic latitude b = -10° - 0° (Fig.7) in 9 pulsars region.

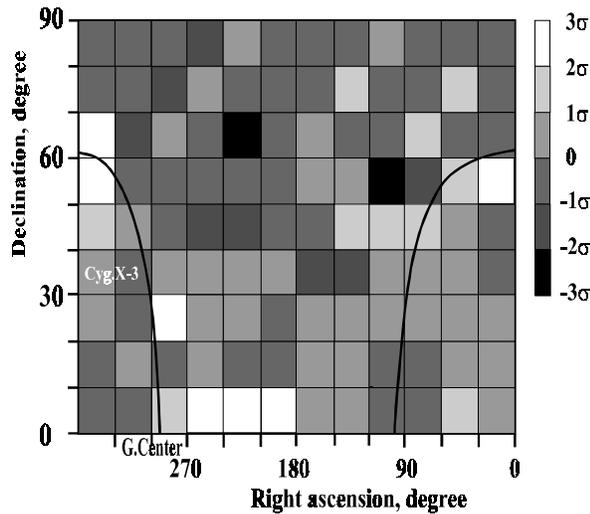

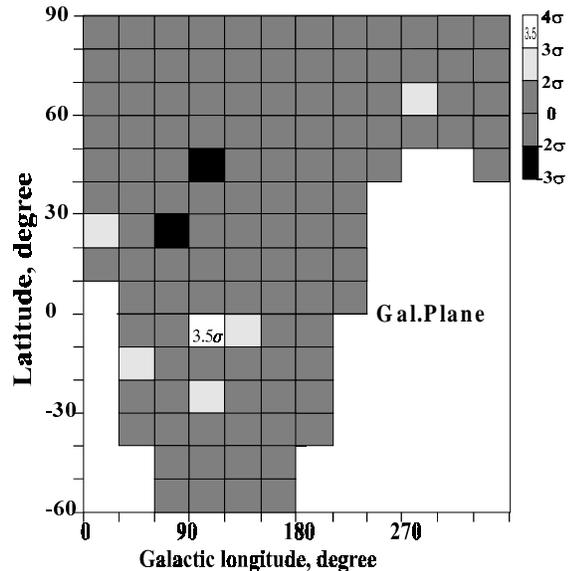

Fig.5. Map of deviations of observed events from expected ones of EAS in equatorial coordinates. Solid lines are the Galactic Plane

Fig.6. Map of deviations of observed events from expected ones of EAS in galactic coordinates

The phase of cosmic ray anisotropy at energy ~ $10^{18}$ eV - $\varphi_1$ ~ 300° in α, $r_1$=4±1%(Hayashida et al.,1998) and $\varphi_1$ ~ 330°(Nesterova,1997), at E~$4.10^{18}$ eV-$\varphi_1$~347°, $r_1$=6.8±1.4% (Mikhailov & Pravdin,1997) can be explained by the particle flux from 9 pulsars with account of their deviation in the magnetic field of the Galaxy(from this side there is the phase $\varphi_1$~1°,$r_1$=16.3±5.9% of considered particles in this paper).

In the model of the large-scale bisymmetric magnetic field of the Galaxy (Rand & Kulkarni, 1989) we determined the deviation angles of the particle arrival from 16 pulsars. We calculated the trajectories of protons with E = $1.3·10^{19}$ eV, which corresponds to the mean particle energy in the considered interval of $(0.4 -4)·10^{19}$ eV. As calculations showed, the deviation angles of protons were <6°. In the case of heavy nuclei (iron) the deviation angles increase up 90°. From here we conclude that the primary particles at r<6° from pulsars are most likely protons, at r>6°- heavy nuclei.

Our calculation show that in galactic latitude at energy ~ $10^{19}$ eV the anisotropy is expected in the case of primary protons and the isotropy - for primary iron nuclei (Fig.6, see also Berezinsky et al,1991, Giller et al.,1994, Lampard et al,1997). On this basis and the observed events in galactic latitude (Fig.7) we conclude that the primary cosmic rays mainly consist of iron nuclei (Fe ~ 80 %). Probably, the pulsars emit both protons and nuclei of iron.

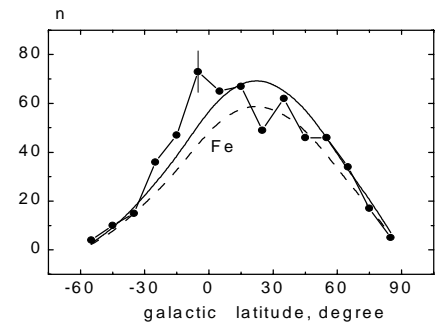

Fig.7.Observed(circles) and expected (solid line) events of EAS in galactic latitude. Dashed line for iron nuclei.

Table. List of 9 pulsars

| Pulsar, PSR | b,  l, degree | dist., kpc |
|---|---|---|
| 0137+57 | -5.0  129.4 | 2.5 |
| 0138+59 | -3.0  129.3 | 3.0 |
| 0144+59 | -2.9  130.1 | 1.1 |
| 0154+61 | -0.6  130.8 | 0.7 |
| 2255+58 | -1.3  108.4 | 4.4 |
| 2319+60 | -0.7  112.0 | 2.8 |
| 2324+60 | -0.9  112.6 | 3.2 |
| 2334+61 | -0.3  114.1 | 1.5 |
| 2351+61 | -0.8  116.1 | 2.5 |

## Summary

The above analysis on search for the relation between the particle arrival directions and pulsars showed that the pulsars of our Galaxy are the sources of ultrahigh energy cosmic rays.

The work is made at the financial support of the Yakutsk EAS array by Russian Foundation for Basic Research (grant N 96-15-96568).